\documentstyle[sprocl,psfig]{article}

\def\be{\begin{equation}}
\def\ee{\end{equation}}
\def\bea{\begin{eqnarray}}
\def\eea{\end{eqnarray}}

\begin{document}

\title{The first fermi in a high energy nuclear collision.}

\author{A. Krasnitz}

\address{UCEH, Universidade do Algarve, Campus de Gambelas, 
P-8000 Faro, Portugal}

\author{Raju Venugopalan}

\address{Physics Department, Brookhaven National Laboratory, Upton, 
NY 11973.}


\maketitle\abstracts{
At very high energies, weak coupling, non--perturbative
methods can be used to study classical gluon production in nuclear
collisions. One observes in numerical simulations 
that after an initial ``formation'' time, the
produced partons are on shell, and their subsequent evolution can be
studied using transport theory. At the initial formation time, a
simple non--perturbative relation exists between the energy and number
densities of the produced partons, and a scale determined by the
saturated parton density in the nucleus.
}

An outstanding problem in high energy scattering is the problem of
initial conditions for particle production~\cite{Levinreview}.  In
perturbative QCD, for processes which involve a hard scale $Q^2 \gg
\Lambda_{QCD}^2$, the hard and soft contributions can be factorized.
The soft contributions are lumped into non--perturbative, process
independent parton distribution functions, while the hard
contributions are computed for each physical process of
interest. For a fixed hard scale of interest $Q^2$, there is
a center of mass energy $\sqrt{s}$ beyond which this approach in
particular, and the operator product expansion (OPE) in general breaks
down~\cite{Mueller1}. However, since the parton densities in this regime are
large, weak coupling classical methods may be applicable~\cite{MV}.  
Wilson renormalization group methods have been developed for this high 
parton density regime~\cite{JKLW}. 

At small $x$, classical parton distributions in a nucleus can be
computed in a model with a dimensionful scale $\mu^2$ proportional
to the gluon density per unit transverse area. In this model, 
parton distributions saturate at a scale $Q_s\propto g^2\mu$. For Au--Au 
collisions at RHIC, one can estimate $Q_s\sim 1$ GeV, and at the LHC, 
the saturation scale will be $Q_s\sim 2$--$3$ GeV. Most of the 
gluons produced therefore have transverse momenta $k_t\sim Q_s$, and 
since this scale for RHIC and LHC is at least marginally a weak coupling 
scale, these classical methods may be applied to study the production and 
initial evolution of partons at RHIC and LHC. 

These classical methods were first applied to nuclear collisions by Kovner, 
McLerran and Weigert~\cite{KLW}. For an interesting alternative approach, see
Ref.~\cite{Balitsky}. 
Assuming boost invariance, and matching the
equations of motion in the forward and backward light cone, they obtained the
following initial conditions for the gauge fields in the $A^\tau =0$ gauge:
$A_\perp^i|_{\tau=0}=A_1^i+A_2^i$\,, and 
$A^{\pm}|_{\tau=0}={\pm}\,{ig\over 2}\,
x^{\pm}[A_1^i,A_2^i]$. Here $A_{1,2}^i(\rho^\pm)$ ($i=1,2$) are the pure gauge 
transverse gauge fields
corresponding to small $x$ modes of incoming nuclei (with light cone sources
$\rho^\pm\,\delta(x^\mp)$) in the $\theta(\pm x^-) \theta(\mp x^+)$
regions respectively of the light cone. 

The sum of two pure gauges in QCD is not a pure gauge--the 
initial conditions therefore give rise to classical gluon radiation in
the forward light cone. For $p_t>>\alpha_S\mu$, the Yang--Mills
equations may be solved perturbatively to quadratic order in
$\alpha_S\mu/p_t$.  After averaging over the Gaussian random sources
of color charge $\rho^\pm$ on the light cone, the perturbative energy
and number distributions of physical gluons were computed by several
authors~\cite{KLW,GM}.  In the small
$x$ limit, it was shown that the classical Yang--Mills result agreed
with the quantum Bremsstrahlung result of Gunion and
Bertsch~\cite{GunionBertsch}.

In Ref.~\cite{AlexRaj1}, we suggested a lattice discretization of the 
classical EFT, suitable for a non--perturbative numerical solution.
Assuming boost
invariance, we showed that in $A^\tau=0$ gauge, the real time
evolution of the small $x$ gauge fields $A_\perp (x_t,\tau), A^\eta
(x_t,\tau)$ is described by the Kogut--Susskind Hamiltonian in
2+1--dimensions coupled to an adjoint scalar field. The lattice
equations of motion for the fields are then determined
straightforwardly by computing the Poisson brackets. The initial
conditions for the evolution are provided by the lattice analogue of
the continuum relations discussed earlier in the text. We impose 
periodic boundary conditions on an $N\times N$ transverse lattice, where 
$N$ denotes the number of sites. 
The physical linear size of the system is $L=a\,N$, where $a$ is the
lattice spacing. It
was shown in Ref.~\cite{AlexRaj2} that numerical computations on a
transverse lattice agreed with lattice perturbation theory at large
transverse momentum.  For details of the numerical procedure, and
other details, we refer the reader to Ref.~\cite{AlexRaj2}.

In our numerical simulations, all the relevant physical information is
compressed in $g^2\mu$ and $L$, and in their dimensionless product
$g^2\mu L$~\cite{RajGavai}.  The strong coupling constant $g$ depends
on the hard scale of interest; $\mu\propto A^{1/6}$ 
depends on the nuclear size, the center of mass energy, and
the hard scale of interest; $L^2$ is the transverse area of the
nucleus. Assuming $g=2$ (or $\alpha_S=1/\pi$), $\mu
=0.5$ GeV ($1.0$ GeV) for RHIC (LHC), and $L=11.6$ fm for
$Au$--nuclei, we find $g^2\mu L\approx 120$ for RHIC and $\approx 240$
for LHC. (The latter number would be smaller for a smaller value of
$g$ at the typical LHC momentum scale.)  As will be discussed later,
these values of $g^2\mu L$ correspond to a region in which one expects
large non--perturbative contributions from a sum to all orders in
$\sim 6\,\alpha_S\mu/p_t$, even if $\alpha_S\ll 1$.  We should mention
here that deviations from lattice perturbation theory, as a function
of increasing $g^2\mu L$, were observed in our earlier
work~\cite{AlexRaj2}.

In Ref.~\cite{latest}, we computed the energy density $\varepsilon$ 
as a function of the proper time $\tau$. This computation on the lattice is 
straightforward. To obtain this result, we computed the 
Hamiltonian density on the lattice for each $\rho^\pm$, and then 
took the Gaussian average (with the weight $\mu^2$) 
over between $40$ $\rho$ trajectories for the larger 
lattices and $160$ $\rho$ trajectories for the smallest ones.

The dependence of $\varepsilon \tau$ as a function of $\tau$ was
investigated in our numerical simulations.  For larger values of
$g^2\mu L$, $\varepsilon\tau$ increases rapidly, develops a transient
peak at $\tau\sim 1/g^2\mu$, and decays exponentially there onwards,
satisfying the relation $\alpha + \beta\,e^{-\gamma\tau}$, to the
asymptotic value $\alpha$ (equal to the lattice $dE/L^2/d\eta$!).
This behavior is satisfied for all $g^2\mu L \ge 8.84$, independently
of $N$.  One can interpret the decay time $\tau_D=1/\gamma/g^2\mu$ as
the appropriate scale controlling the formation of gluons with a
physically well defined energy. In other words, $\tau_D$ is the
``formation time''in the sense used by Bjorken~\cite{Bj}. 

\begin{figure}[h]
\vspace{-5cm}
\centerline{\hspace{-35cm}{\hbox{\psfig{figure=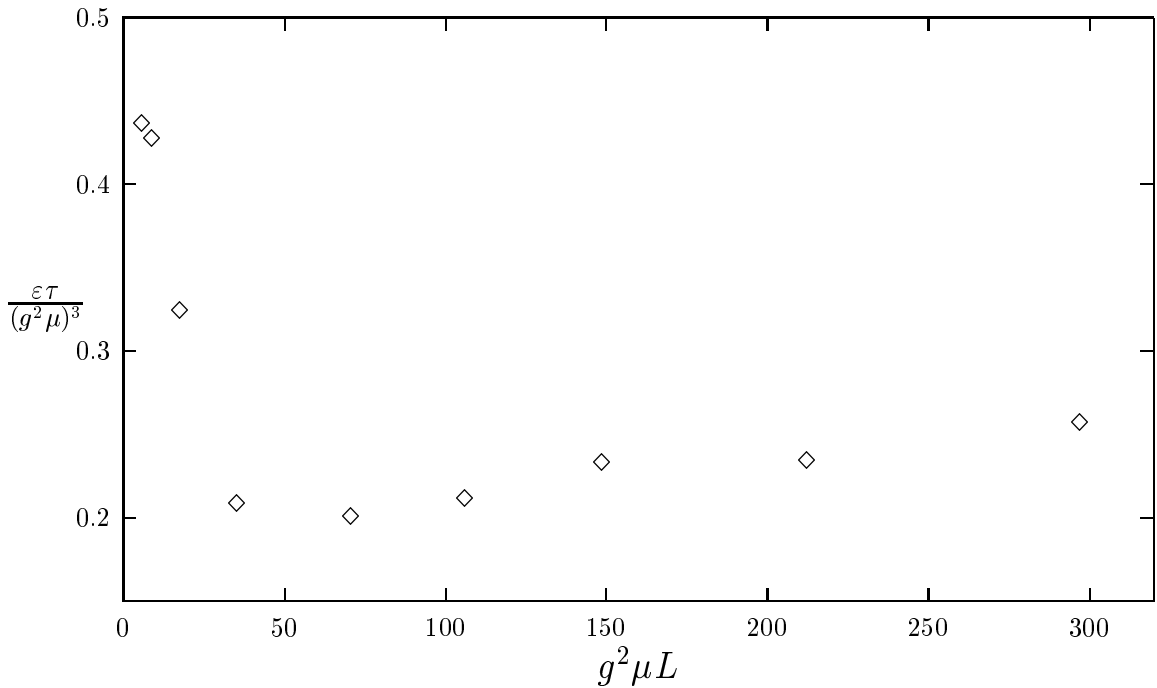,
width=12cm}}}}
\vskip -14cm
\caption{$\varepsilon\tau/(g^2\mu)^3$ extrapolated to the continuum limit: 
$f$ as a function of $g^2\mu L$. The error bars are smaller than the 
plotting symbols.}
\label{eXtvsmuL}
\end{figure}
The physical energy per unit area per unit rapidity of 
produced gluons can be defined in terms of a function $f(g^2\mu L)$ as
\be
{1\over L^2}\, {dE\over d\eta} = {1\over g^2}\,f(g^2\mu L)\,(g^2\mu)^3 \, .
\label{energydensity}
\ee
The function $f$ here is obtained by extrapolating our results for finite 
lattice spacings to the continuum limit. In the region of physical interest 
for heavy ion collisions, $f$ varies very slowly. It changes by $\sim 25$\% 
for nearly an order of magnitude change in $g^2\mu L$. The saturation 
scale $Q_s\sim 6\, \alpha_S\mu$--one can therefore re--write our result for 
the energy density in terms of $Q_s$.

Doing so, we confirmed that our results are consistent with an estimate by
A. H. Mueller~\cite{Muell2} for the number of produced gluons per unit
area per unit rapidity. He obtains $dN/L^2/d\eta =
c\,(N_c^2-1)\,Q_s^2/4\pi^2 \,\alpha_S\,N_c$, and argues that the
number $c$ is a non--perturbative constant of order unity. If most of
the gluons have $p_t\sim Q_s$, then $dE/L^2/d\eta =
c^\prime\,(N_c^2-1)\,Q_s^3/4\pi^2\,\alpha_S\,N_c$ which is of the
same form as our Eq.~\ref{energydensity}.  In the $g^2\mu L$ region of
interest, our function $f\approx 0.23$--$0.26$. 
We obtain $c^\prime = 4.3$--$4.9$. 
Since one expects a distribution in momenta about $Q_s$, it is very likely that
$c^\prime$ is at least a factor of $2$ greater than $c$--thereby
yielding a number of order unity for $c$ as estimated by Mueller. This
coefficient can be determined more precisely when we compute the
non--perturbative number and energy distributions~\cite{AlexRaj3}.

In Ref.~\cite{latest}, we estimated
the initial energy per unit rapidity of produced
gluons at RHIC and LHC energies. We did so by extrapolating from our
SU(2) results to SU(3) assuming the $N_c$ dependence to be
$(N_c^2-1)/N_c$ as in Mueller's formula. At late times, the energy
density is $\varepsilon = (g^2\mu)^4\,f(g^2\mu L)\,\gamma(g^2\mu
L)/g^2$, where the formation time is $\tau_D=1/\gamma(g^2\mu
L)/g^2\mu$ as discussed earlier. We find  $\varepsilon^{RHIC}\approx
66.49$ GeV/fm$^3$ and $\varepsilon^{LHC}\approx 1315.56$
GeV/fm$^3$. Multiplying these numbers by the initial volumes at the
formation time $\tau_D$, we obtained the classical Yang--Mills estimate
for the initial energies per unit rapidity $E_T$ to be $E_T^{RHIC}\approx
2703$ GeV and $E_T^{LHC}\approx 24572$ GeV respectively.

Compare these numbers to results presented recently by
Kajantie~\cite{Keijo} for the mini--jet energy (computed for 
$p_t > p_{sat}$, where $p_{sat}$ is a saturation scale akin to $Q_s$) in 
the pQCD mini--jet approach~\cite{minijets}. He
obtains $E_T^{RHIC} = 2500$ GeV and $E_T^{LHC}=12000$. The remarkable
closeness between our results for RHIC is very likely a
coincidence. Kajantie's result includes a $K$ factor of $1.5$--estimates 
range from $1.5$--$2.5$~\cite{KeijoKari}. For the latest estimates from our
Finnish colleagues, see the preprint of Eskola et al.~\cite{Karietal}. 
If we pick a recent value of 
$K\approx 2$~\cite{Andrei}, we obtain as 
our final estimate, $E_T^{RHIC}\approx 5406$ GeV and $E_T^{LHC}\approx 
49144$ GeV.

We can also boldly estimate the number of produced gluons at central 
rapidities. As mentioned in the preceding text, the value of the 
constant $c$ in the expression for the number distribution is currently being 
computed numerically. We obtain for Au--Au 
collisions in one unit of rapidity the result that $N_{RHIC} = 714\cdot c$ and 
$N_{LHC}= 2855\cdot c$. Given that the corresponding constant for the 
energy density was larger, we would anticipate that it is more likely that 
$c=2$--$2.5$. Taking the higher value, we obtain $N_{RHIC} = 1785$ and 
$N_{LHC}=7138$. Again, these values are very close to those of Eskola et al.
~\cite{Karietal}. Note that they consider Pb--Pb collisions and their results 
include a $K$ factor of 2. The purpose of this simple exercise is primarily 
to confirm that our numbers are not wildly divergent from mini--jet 
calculations. Our results are typically a factor of two larger 
(more at LHC) but this is 
easily understood because our results include all momentum modes. 

The number density of these out--of--equilibrium gluons can be related to the 
equilibrium entropy: $S_{glue} = 3.6\cdot N_{glue}$. This is particularly so 
at the LHC, where because $Q_s^2\gg \Lambda_{QCD}^2$, elastic scattering 
dominates. The equilibrium entropy of gluons is, to within a factor of two, 
(which can be quantified in one's thermal+hydro model of choice), the 
entropy of pions.

\section*{Acknowledgments}
I would like to thank Dr. A. Dumitru for a useful discussion. This research 
was supported by DOE Nuclear Theory at BNL.

\end{document}